\documentclass[%
reprint,
superscriptaddress,
amsmath,
amssymb,
aps,
pra
]{revtex4-1}

\usepackage{graphicx}
\usepackage{dcolumn}
\usepackage{xcolor}
\usepackage[colorlinks=true,allcolors=blue]{hyperref}
\usepackage[all]{hypcap}

\begin{document}
\title{Ground-state quantum geometry in superconductor--quantum dot chains}

\author{R.~L.~Klees}
\affiliation{Fachbereich Physik, Universit{\"a}t Konstanz, D-78457 Konstanz, Germany}
\author{J.~C.~Cuevas}
\affiliation{Departamento de F\'{i}sica Te\'{o}rica de la Materia Condensada and Condensed Matter Physics Center (IFIMAC), Universidad Aut\'{o}noma de Madrid, E-28049 Madrid, Spain}
\author{W.~Belzig}
\affiliation{Fachbereich Physik, Universit{\"a}t Konstanz, D-78457 Konstanz, Germany}
\author{G.~Rastelli}
\affiliation{Fachbereich Physik, Universit{\"a}t Konstanz, D-78457 Konstanz, Germany}
\affiliation{Zukunftskolleg, Universit{\"a}t Konstanz, D-78457 Konstanz, Germany}
\affiliation{INO-CNR BEC Center and Dipartimento di Fisica, Universit{\`a} di Trento, I-38123 Povo, Italy}

\date{\today}

\begin{abstract}
	Multiterminal Josephson junctions constitute engineered topological systems in arbitrary synthetic dimensions defined by the superconducting phases. 
	Microwave spectroscopy enables the measurement of the quantum geometric tensor, a fundamental quantity describing both the quantum geometry and the topology of the emergent Andreev bound states in a unified manner.
	In this work we propose an experimentally feasible and scalable multiterminal setup of $N$ quantum dots connected to $N+1$ superconducting leads which allows to deterministically study nontrivial topology in terms of the Chern number of the noninteracting ground state. 
	An important result is that the nontrivial topology in a linear chain appears beyond a threshold value of the nonlocal proximity-induced pairing potential which represents the novel theoretical key ingredient of our proposal. 
	Moreover, we generalize the microwave spectroscopy scheme to the multiband case and show that the elements of the quantum geometric tensor of the noninteracting ground state can be experimentally accessed from the measurable oscillator strengths at low temperature.
\end{abstract}

\maketitle

\section{\label{sec:introduction}Introduction}
The concept of topology plays an important role in modern branches of theoretical physics by explaining the quantized nature of certain physical phenomena.
The pursue to find novel types of topologically nontrivial systems providing quantized observables classified by integers has created a new research field called topological band theory \cite{Bansil:2016}. 
Since then, a huge variety of topological quantum matter has been predicted and discovered such as topological insulators \cite{Kane:2005gb,Hasan:2010ku}, topological semimetals \cite{Armitage:2018dg}, topological superconductors \cite{Sato:2017go}, as well as non-Hermitian (e.g., open/dissipative) topological systems \cite{Kawabata:2019}.
Out of these, the superconducting systems are also promising platforms for topologically protected quantum computation that relies on non-Abelian exchange statistics, i.e., braiding, of Majorana zero modes appearing at the Fermi energy \cite{Kitaev:2001,Kitaev:2003,Fu:2008} and their appearance was recently verified experimentally \cite{Yang:2019}. 
Furthermore, qubits based on the Andreev bound states (ABS) appearing inside a Josephson junction have also been proposed \cite{Chtchelkatchev:2003,Zazunov:2003}. 
These bound states can be experimentally accessed and coherently manipulated by microwave \cite{Bretheau:2013,Janvier:2015,vanWoerkom:2017,Hays:2018}, tunneling \cite{Pillet:2010}, and supercurrent spectroscopy \cite{Bretheau:2013bt} if the junction is embedded in an rf superconducting quantum interference device (SQUID).

More recently, multiterminal Josephson junctions (MJJs) consisting of many superconducting terminals have been theoretically investigated and shown to exhibit topologically nontrivial physics
\cite{vanHeck:2014,Padurariu:2015,Yokoyama:2015,Riwar:2016,Yokoyama:2017,Eriksson:2017,Xie:2017,Meyer:2017,Xie:2018,Repin:2019,Gavensky:2019,Houzet:2019,Trif:2019,Klees:2020,Chen:2020b,Sakurai:2020,Repin2020}.
In such multiterminal systems, topology emerges in the synthetic space of superconducting phases and the integer-valued Chern number can manifest itself in a quantized transconductance between two terminals \cite{Riwar:2016,Eriksson:2017,Xie:2018,Repin:2019,Xie:2017,Meyer:2017,Gavensky:2019,Houzet:2019,Sakurai:2020}.
The advantage of these systems is that, in principle, an arbitrary number of synthetic dimensions can be implemented by simply increasing the number of superconducting leads and that building blocks can be conventional materials, although also topological superconductors hosting Majorana zero modes have been studied in this context \cite{Gavensky:2019,Houzet:2019,Trif:2019,Sakurai:2020}.
Moreover, it has been recently suggested to use microwave spectroscopy to measure the more fundamental quantum geometric tensor of ABS, which provides both the information about the geometry of the state manifold and the topological information contained in the Berry curvature \cite{Klees:2020}.
While the Chern number follows directly from the local Berry curvature by a simple integration, the local metric tensor carries important information on, for instance, energy fluctuations and the noise spectral function \cite{Kolodrubetz:2013bf,Neupert:2013eu}, as well as quantum phase transitions \cite{Zanardi:2007}, wave packet dynamics \cite{Bleu:2018,Solnyshkov:2020}, the superfluid weight \cite{Peotta2015,Julku2016,Liang2017}, energy corrections of excitons \cite{Srivastava2015}, the orbital magnetic susceptibility \cite{Gao:2014,Gao:2015,Piechon:2016}, and magnetic exchange constants \cite{Freimuth2017}.
Finally, it is also possible to strongly drive topologically trivial Josephson systems to eventually generate nontrivial (Floquet) topology \cite{Gavensky:2018em,Venitucci:2018gb}.

\begin{figure*}
	\centering
	\includegraphics[width=\linewidth]{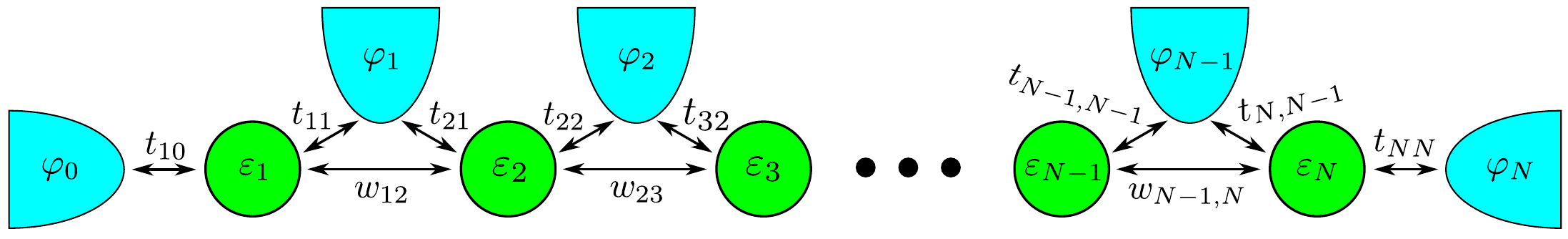}
	\caption{Model of the superconductor-quantum dot chain. The chain consists of $N$ quantum dots (each with onsite energy $\varepsilon_j$, $j = 1,\ldots,N$) which are connected to $N+1$ $s$-wave superconducting leads (each with a pairing phase $\varphi_k$, $k = 0,\ldots,N$). The couplings between the quantum dots are $w_{l,l+1}$ ($l = 1,\ldots,N-1$), while the  couplings between the quantum dots and the superconducting leads are $t_{j,j-1}$ and $t_{jj}$.}
	\label{fig:fig1}
\end{figure*}
So far, many experiments to measure quantum geometry (i.e., the Berry curvature and/or the quantum metric) and the related topology have already been successfully carried out in, e.g., ultracold fermionic atoms \cite{Flaeschner:2016,Asteria:2019}, superconducting qubits and qutrits \cite{Schroer:2014,Tan:2018}, superconducting transmon qubits \cite{Tan:2019}, superconducting quantum circuits \cite{Roushan:2014}, qubits in NV centers in diamond \cite{Yu:2019}, high-finesse planar microcavities \cite{Gianfrate:2020}, and coupled fiber loops \cite{Wimmer:2017}, the last two representing examples of the emerging field of topological photonics \cite{Ozawa:2019}.

Additional interesting directions are topological fermion-parity \cite{Kotetes:2019} and Cooper-pair pumps \cite{Geerligs:1991,Pekola:1999,Aunola:2003,Leone:2008,Erdman:2019}, as well as Josephson tunnel circuits \cite{2020:Fatemi,Peyruchat:2020,Herrig2020}, in which Weyl singularities in parameter space are responsible for a quantized adiabatic transfer of single charges (charge $1e$) and Cooper pairs (charge $2e$).

Although recently a diffusive three-terminal MJJ in a double-SQUID configuration has been both experimentally and theoretically investigated on its topological real-space properties  \cite{Strambini:2016gt} followed by a detailed theoretical study \cite{Amundsen:2017}, it is still challenging to create MJJs \cite{Pillet:2010,Plissard:2013kk}.
Furthermore, despite the fact that first experiments towards ballistic MJJs have been performed \cite{Draelos:2019gl,Pankratova:2020,Arnault2020}, the more general theoretical proposals of MJJs discussed above are expected to be quite challenging to realize since the necessary space for experimental control gates decreases as the number of superconducting terminals increases.
Such circular arrangements of the terminals will eventually have an adverse effect on the scalability of these systems.

In this work, we propose a linear design of an MJJ as an alternative and scalable device to enable the deterministic study quantum geometry and topology in topological Josephson matter.
This setup consists of a chain of quantum dots which are coupled to superconducting leads, as presented in Fig.~\ref{fig:fig1}.
We think that this system is easier to realize and manipulate in experiments since there is enough space from below for additional necessary control gates and it could be implemented with already available ingredients such as carbon nanotubes or semiconducting nanowires contacted by superconductors \cite{Pillet:2010,LevyYeyati:1997,Buitelaar:2003,Hofstetter:2009,Herrmann:2010,Fulop:2015}.
Moreover, we expect that it will be straightforward to extend the length of the chain and, hence, increase the number of dots and leads at will.
We will show that such a system allows for the emergence of topologically nontrivial ABS and, therefore, represents an ideal platform to experimentally study quantum geometry and topology in a deterministic way, in contrast to prior proposals based on statistical methods \cite{Riwar:2016,Eriksson:2017,Meyer:2017}. 
We will also show how to apply microwave spectroscopy in such multiband systems to measure the quantum geometry of the noninteracting ground state (GS) as a whole, which is important for (quantized) quantum transport phenomena due to the collection of occupied single-particle states.

The rest of the paper is organized as follows: 
In Sec.~\ref{sec:model}, we introduce and investigate the effective low-energy Hamiltonian of a linear chain of $N$ dots coupled to $N+1$ superconductors. 
We also define the topological invariant of the noninteracting GS, i.e., the Chern number, and analyze the effects of the various parameters of this model on the topological phases of this system. 
In Sec.~\ref{sec:geometryAndTopology}, we discuss the application of microwave spectroscopy in such systems.
Importantly, we find that, although the quantum geometry of the individual ABS is inaccessible at zero temperature, the topology and the quantum geometry of the noninteracting GS formed by many occupied fermionic states can be accessed even in the multiband case.
We discuss our findings and conclude our work in Sec.~\ref{sec:discussion}.
Further details of derivations can be found in the Appendixes~\ref{app1:effHam}, \ref{app3:manybody}, and \ref{app2:PMS}.
\section{\label{sec:model}Effective low-energy model and topology}
As described above, we start by considering a linear chain of $N+1$ BCS superconducting leads connected to $N$ quantum dots as shown in Fig.~\ref{fig:fig1}.
Each of the quantum dots is spin-degenerate with an onsite energy $\varepsilon_j$ $(j = 1,\ldots,N)$ and each $s$-wave superconductor has a pairing potential $\Delta_k$ with a pairing phase $\varphi_k$ $(k = 0 , \ldots , N)$.
The dot $j$ is coupled to two neighboring superconducting leads $j-1$ and $j$ via the hoppings $t_{j,j-1}$ and $t_{jj}$, respectively.
To be more general, we include the effect of a direct tunneling between the  dots without passing through the superconductors, which is described by a direct coupling $w_{l,l+1}$ ($l = 1, \ldots, N-1$) between the dots $l$ and $l+1$.
By integrating out the subsystem of the superconducting leads and by considering the limit of large superconducting gaps $\Delta_k \to \infty$, we obtain an effective low-energy Hamiltonian describing the quantum dots proximity coupled to the superconductors.
The derivation of the effective Hamiltonian of our model is presented in Appendix~\ref{app1:effHam}.

The resulting Hamiltonian has the form $H_{\mathrm{eff}} = \boldsymbol{d}^\dag \hat{H}_{\mathrm{eff}}^{(N)} \boldsymbol{d}$, 
where the $2N \times 2N$ block-tridiagonal matrix Hamiltonian is given by
\begin{align}
	\hat{H}_{\mathrm{eff}}^{(N)} = \begin{pmatrix}
		\hat{H}_1 & \hat{V}_{12} & 0 & \cdots & 0 & 0 \\
		\hat{V}_{12} & \hat{H}_{2} & \hat{V}_{23} & \cdots & 0 & 0 \\
		0 & \hat{V}_{23} & \hat{H}_{3} & \cdots & 0 & 0 \\
		\vdots & \vdots  & \vdots & \ddots & \vdots  & \vdots \\
		0 & 0  & 0 & \cdots & \hat{H}_{N-1} & \hat{V}_{N-1,N} \\
		0 & 0  & 0 & \cdots & \hat{V}_{N-1,N} & \hat{H}_N \\
	\end{pmatrix},
	\label{eq:eq1}
\end{align}
which is written in the basis defined by the  $2N$-dimensional Nambu spinor $\boldsymbol{d}^{\dag} = (d_{1\uparrow}^{\dag},d_{1\downarrow}^{\phantom\dag}, d_{2\uparrow}^{\dag},d_{2\downarrow}^{\phantom\dag}, \ldots, d_{N\uparrow}^{\dag},d_{N\downarrow}^{\phantom\dag})$, with $d_{j\sigma}^{(\dag)}$ being the annihilation (creation) operator of an electron with spin $\sigma \in \{\uparrow,\downarrow\}$ on the $j$-th dot. 
The $2\times 2$ Nambu blocks are given by
\begin{align}
	\hat{H}_j &= \begin{pmatrix}
		\varepsilon_j & z_j \\
		z_j^* & -\varepsilon_j
	\end{pmatrix},
	\
	\hat{V}_{l,l+1} = \begin{pmatrix}
		w_{l,l+1} & z_{l,l+1} \\
		z_{l,l+1}^* & -w_{l,l+1}
	\end{pmatrix} ,
\end{align}
where the effective local and nonlocal proximity-induced pairing potentials on the dots are defined as
\begin{subequations}
\begin{align}
	z_j &= \Gamma_{j}^{(j-1)} e^{i \varphi_{j-1}} + \Gamma_{j}^{(j)} e^{i \varphi_j} , \\
	z_{l,l+1} &= \Gamma_{l,l+1}^{(l)} e^{i \varphi_l} ,
\end{align}
\end{subequations}
with $\Gamma_{j}^{(k)} = \pi N_0 t_{jk}^2$ and $\Gamma_{l,l+1}^{(l)} = \pi N_0 p_l t_{ll} t_{l+1,l}$, respectively, where $N_0$ is the normal density of states at the Fermi energy.
In passing by, we modified the nonlocal pairings by dimensionless factors $p_l \in [0,1]$ motivated by experiments in which typically $\Gamma_{l,l+1}^{(l)} < \sqrt{ \Gamma_{l}^{(l)} \Gamma_{l+1}^{(l)} }$ \cite{Hofstetter:2009,Herrmann:2010}.
This modification allows us to treat the nonlocal couplings as independent parameters from the local ones.
In general, the parameters $p_l$ modifying the nonlocal couplings $\Gamma_{l,l+1}^{(l)}$ depend on the geometrical details of the contacts as well as the coherence length of the Cooper pairs \cite{Recher:2001,Feinberg:2003}.
\begin{figure*}
	\centering
	\includegraphics[width=\linewidth]{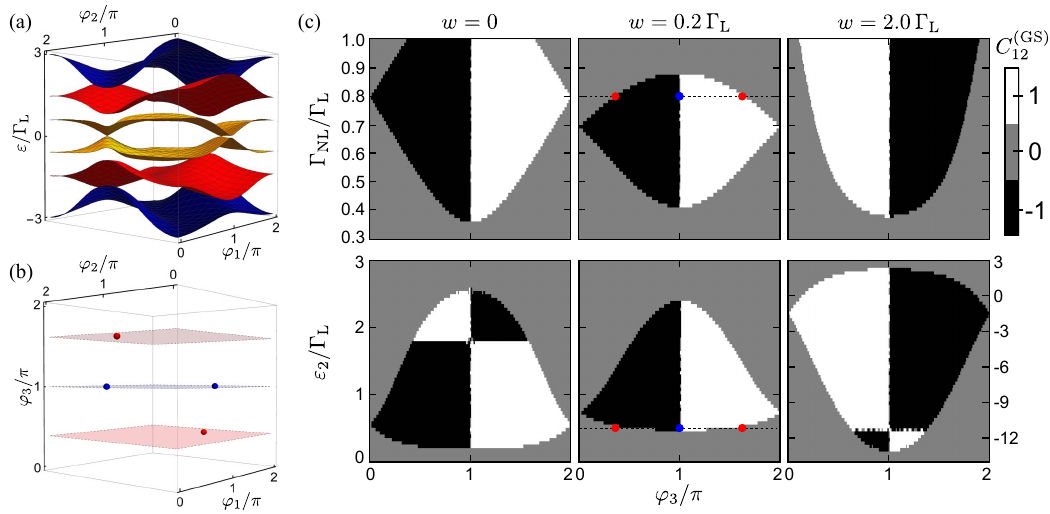}
	\caption{
		(a) ABS energies $\varepsilon_{\mathrm{A},-3} , \ldots , \varepsilon_{\mathrm{A},3}$ (from bottom to top) for $N=3$ dots. Energy bands with the same color represent PH-symmetric ABS with energies $\varepsilon_{\mathrm{A},\pm j}$. Parameters: $\varepsilon_1 = -\varepsilon_2 = \varepsilon_3 = -\Gamma_\mathrm{L}/2$, $\varphi_0 = 0$, $\varphi_3 = \pi$, $w = 0.2 \, \Gamma_\mathrm{L}$, $\Gamma_\mathrm{NL} = 0.8 \, \Gamma_\mathrm{L}$.
		(b) Locations of the four Weyl points $\boldsymbol{\varphi}_{\mathrm{W}} \in \mathbb{T}^3$ for which $\varepsilon_{\mathrm{A},\pm 1}(\boldsymbol{\varphi}_{\mathrm{W}}) = 0$ for the same parameters as in panel (a). The color indicates the topological charge of the Weyl points. The two blue Weyl points in the $\varphi_3 = \pi$ plane are visible as zero-energy band crossings in panel (a).
		(c) Topological phase diagrams of the ground state Chern number $C_{12}^{(\mathrm{GS})}$ for $N = 3$ dots for different values of the direct tunneling $w$ between the dots (columns) as a function of the phase $\varphi_3$. The upper row shows the phase diagram as a function of the nonlocal coupling $\Gamma_\mathrm{NL}$ for fixed $\varepsilon_{2} = \Gamma_\mathrm{L}/2$, while the lower row shows it as a function of the central dot level $\varepsilon_{2}$ for fixed $\Gamma_\mathrm{NL} = 0.8 \, \Gamma_\mathrm{L}$. All other parameters are the same as in panel (a). The dashed line with blue and red dots in the phase diagrams for $w = 0.2 \, \Gamma_\mathrm{L}$ mark the locations of the planes with fixed $\varphi_3$ in which the Weyl points of the same color appear in panel (b).}
	\label{fig:fig2}
\end{figure*}

Finally, as in a system of $N+1$ superconducting leads only $N$ superconducting phases are independent, gauge-invariance allows us to set one of the superconducting phases to zero.
We will consider $\varphi_0 = 0$ in the following as a reference and the remaining $N$ superconducting phases $\boldsymbol{\varphi} = (\varphi_1 , \ldots , \varphi_N) \in [0,2\pi)^N$ define an $N$-dimensional periodic compact space $\mathbb{T}^N$ ($N$-torus) in analogy to the quasimomenta in a periodic crystal with a first Brillouin zone.

For $N$ quantum dots coupled to $N+1$ superconducting leads in the low-energy limit, there are $2N$ ABS.
Defining the sets of labels $D^{\pm} = \{\pm 1 , \ldots, \pm N\}$ for occupied ($D^-$) and empty ($D^+$) states at zero temperature, the $2N$ ABS $|\psi_n\rangle$ ($n \in D^- \cup D^+)$ satisfy $\hat{H}_{\mathrm{eff}}^{(N)} |\psi_n\rangle = \varepsilon_{\mathrm{A},n} |\psi_n\rangle$, where $\varepsilon_{\mathrm{A},n}$ is the energy of the $n$-th ABS.
In the following, we will consider the ABS being ordered as $\varepsilon_{\mathrm{A},-N} \leq \ldots \leq \varepsilon_{\mathrm{A},-1} \leq 0 \leq \varepsilon_{\mathrm{A},1} \leq \ldots \leq \varepsilon_{\mathrm{A},N}$ without loss of generality.

The $2N$ ABS come in $N$ particle-hole (PH) symmetric pairs and fulfill $|\psi_{-j}\rangle = P^{(N)} |\psi_{j}\rangle$ $(j \in D^+)$, where $P^{(N)} = i (\openone^{(N)} \otimes \tau_2) K$ is the operator of PH symmetry, $K$ is complex conjugation, $\openone^{(N)}$ is the $N\times N$ identity matrix, and $\tau_2$ is the second Pauli matrix in Nambu space.
Furthermore, the energies of a PH-symmetric pair are related by $\varepsilon_{\mathrm{A},-j} = -\varepsilon_{\mathrm{A},j}$ since $P^{(N)}$ satisfies the anticommutation relation $\{\hat{H}_{\mathrm{eff}}^{(N)}, P^{(N)}\} = 0$.

Due to the large number of parameters, we choose equal local couplings $\Gamma_{j}^{(k)} = \Gamma_\mathrm{L}$, equal nonlocal couplings $\Gamma_{l,l+1}^{(l)} = \Gamma_\mathrm{NL}$, and equal interdot couplings $w_{l,l+1} = w$ for simplicity. 
As an example, we show the ABS for $N=3$ dots in Fig.~\hyperref[fig:fig2]{2(a)}, which also shows that this system allows for the appearance of Weyl singularities, i.e., points of degeneracy $\boldsymbol{\varphi}_\mathrm{W} \in \mathbb{T}^3$ at which two energy bands cross [Fig.~\hyperref[fig:fig2]{2(b)}].
In this case, we have set the energies of the quantum dots equal in magnitude with an alternating sign.
However, Weyl singularities will also appear in more general situations.
As we will see below, Weyl points for which $\varepsilon_{\mathrm{A},\pm 1}(\boldsymbol{\varphi}_\mathrm{W}) = 0$ are responsible for topological phase transitions between regions of different values of the Chern number of the GS of the system.

The superconducting phases $\boldsymbol{\varphi}$ define $N$ synthetic $U(1)$ gauge fields for which we define gauge connection 1-forms $\mathcal{A}^{(n)} = \sum_{\alpha=1}^N a_\alpha^{(n)} \mathrm{d}\varphi_{\alpha}$ for each of the ABS $|\psi_{n}\rangle$, where $a_\alpha^{(n)} = i \langle\psi_{n}|\partial_\alpha\psi_{n}\rangle$ is the (Abelian) Berry connection \cite{Berry:1984} and $\partial_{\alpha} \equiv \partial/\partial \varphi_{\alpha}$.
The gauge-invariant curvature 2-form $\mathcal{F}^{(n)}$ of the $n$-th single-particle state is defined as the exterior derivative of $\mathcal{A}^{(n)}$, i.e., $\mathcal{F}^{(n)} = \mathrm{d}\mathcal{A}^{(n)} = (1/2) \sum_{\alpha,\beta=1}^N f_{\alpha\beta}^{(n)} \mathrm{d}\varphi_{\alpha} \wedge \mathrm{d}\varphi_{\beta}$, where $f_{\alpha\beta}^{(n)} = \partial_{\alpha} a_\beta^{(n)} - \partial_{\beta} a_\alpha^{(n)}$ is the (Abelian) Berry curvature \cite{Nakahara:2003}.
Finally, the first Chern numbers of an ABS $|\psi_{n}\rangle$ are calculated as
\begin{align}
	\label{eq:chernSingleParticle}
	c_{\alpha\beta}^{(n)} = \frac{1}{2\pi} \int_{0}^{2\pi} \int_{0}^{2\pi}  f_{\alpha\beta}^{(n)} \, \mathrm{d}\varphi_{\alpha} \, \mathrm{d}\varphi_{\beta} \, ,
\end{align}
which are defined by an integration in the $(\varphi_{\alpha},\varphi_{\beta})$ subplane of the $N$-torus $\mathbb{T}^N$.
Note that, due to PH symmetry, we have $a_\alpha^{(-j)} = -a_\alpha^{(j)}$ and, consequently, $f_{\alpha\beta}^{(-j)} = -f_{\alpha\beta}^{(j)}$ and $c_{\alpha\beta}^{(-j)} = -c_{\alpha\beta}^{(j)}$.
We use the numerical algorithm developed by Fukui, Hatsugai, and Suzuki in Ref.~\cite{Fukui:2005er} for a stable and gauge-independent calculation of the Chern numbers.

While the topological properties of an ABS $|\psi_{n}\rangle$ in terms of the Chern numbers $c_{\alpha\beta}^{(n)}$ is encoded in its local Berry curvature $f_{\alpha\beta}^{(n)}$, the full quantum geometry of each of these ABS is described by the local quantum geometric tensor (QGT) \cite{Kolodrubetz:2017jg}
\begin{align}
	q^{(n)}_{\alpha\beta} &= 
	\langle \partial_\alpha \psi_{n} |( 1 - |\psi_{n} \rangle\langle\psi_{n} |) |\partial_\beta\psi_{n}\rangle \, .
	\label{eq:QGTnew}
\end{align}
In particular, we obtain the (Fubini-Study) metric tensor $g^{(n)}_{\alpha\beta} = \mathrm{Re}(q^{(n)}_{\alpha\beta})$ from the real part of the QGT that provides a measure of distance $\mathrm{d}s^2 = \sum_{\alpha,\beta=1}^N g^{(n)}_{\alpha\beta}  \mathrm{d}\varphi_{\alpha} \mathrm{d}\varphi_{\beta}$ on the $n$-th ABS manifold of physical states, while the Berry curvature $f^{(n)}_{\alpha\beta} = -2 \, \mathrm{Im}(q^{(n)}_{\alpha\beta})$ is obtained from the imaginary part of the QGT that describes the geometric phase of a physical quantum state.

The noninteracting GS of the system is defined by the collection of all occupied single-particle states below the Fermi energy, i.e., the $N$ ABS with energies $\varepsilon_{\mathrm{A},j} \leq 0$ for $j \in D^{-}$.
As shown in Appendix~\ref{app3:manybody}, the metric tensor and the Berry curvature of the noninteracting GS can be constructed from the single-particle contributions as
\begin{subequations}
	\label{eq:groundStateQGT}
	\begin{align}
		G^{\mathrm{(GS)}}_{\alpha\beta} 
		&= \sum_{j \in D^-} g_{\alpha\beta}^{(j)} 
		- \sum_{j,k \in D^- \atop j \neq k} 
		\langle \partial_{\alpha} \psi_j | \psi_k \rangle
		\langle \psi_k | \partial_{\beta} \psi_j \rangle \, ,
		\\
		\label{eq:berryGSmany}
		F^{\mathrm{(GS)}}_{\alpha\beta} 
		&= \sum_{j \in D^-} f_{\alpha\beta}^{(j)} \, .
	\end{align}
\end{subequations}
While the result for the Berry curvature $F^{\mathrm{(GS)}}_{\alpha\beta}$ is already well-known, the counter-intuitive result for the metric tensor $G^{\mathrm{(GS)}}_{\alpha\beta}$ shows that it is not a simple sum over the metric tensors of the individual single-particle states.

As a direct consequence of Eqs.~\eqref{eq:chernSingleParticle} and \eqref{eq:berryGSmany}, the Chern number of the noninteracting GS is simply given by the sum of all individual Chern numbers of the occupied single-particle states, i.e.,
\begin{align}
	C_{\alpha\beta}^{\mathrm{(GS)}} = \sum_{j \in D^-} c_{\alpha\beta}^{(j)} .
	\label{eq:chernGS}
\end{align}
Note that $C_{\alpha\beta}^{\mathrm{(GS)}}$ only changes if there is a zero-energy crossing between the two PH-symmetric ABS $|\psi_{-1}\rangle$ and $|\psi_{1}\rangle$.
Although there are of course Weyl points for which other energy bands cross changing the individual Chern numbers from, e.g., 0 to $\pm1$, the sum of them entering $C_{\alpha\beta}^{\mathrm{(GS)}}$ in Eq.~\eqref{eq:chernGS} will still be unchanged.

In order to obtain topologically nontrivial states in terms of $C_{\alpha\beta}^{\mathrm{(GS)}}$, the minimal number of quantum dots in this system is $N = 3$ such that there are three tunable phase differences, as has already been discussed in Refs.~\cite{Yokoyama:2015,Riwar:2016,Eriksson:2017,Xie:2018,Repin:2019,Klees:2020}.
To analyze and discuss the stability of the topological phases for this case, we show topological phase diagrams in Fig.~\hyperref[fig:fig2]{2(c)} in terms of the GS Chern number $C_{12}^{\mathrm{(GS)}}$.
We see that there are large and stable regions for which nontrivial regions of $C_{12}^{\mathrm{(GS)}} \neq 0$ appear.
This stability is due to the separation of Weyl points $\boldsymbol{\varphi}_\mathrm{W}$ of different topological charge [Fig.~\hyperref[fig:fig2]{2(b)}].
These Weyl points move continuously in $\mathbb{T}^3$ while the parameters of the system are changed, making these regions robust against small fluctuations.
As long as two Weyl points of different topological charge do not meet and annihilate, topological regions will be present.

As shown in Fig.~\hyperref[fig:fig2]{2(c)}, the first important result is that the nontrivial topology in a linear chain appears only in the presence of a nonlocal pairing potential which represents the novel theoretical key ingredient of our proposal. 
A nonzero Chern number appears only beyond a certain threshold value of the nonlocal pairing coupling $\Gamma_\mathrm{NL}$.
Interestingly, we see that it is not necessary that the quantum dots are directly coupled ($w=0$) for the Chern number to be nontrivial.
For a small coupling $w=0.2\, \Gamma_\mathrm{L}$ we still find topologically nontrivial regions. 
Remarkably, a larger interdot coupling of $w=2 \,  \Gamma_\mathrm{L}$  can also enhance the size of topological regions in parameter space, although the two regions of finite Chern number are interchanged.
However, if the interdot coupling becomes too strong ($w \gg \Gamma_\mathrm{L}$, not shown), topological regions completely disappear since the effect of the superconductors will be negligible. 
In addition, we see that the nonlocal couplings $\Gamma_\mathrm{NL}$ can be considerably smaller than the local couplings $\Gamma_\mathrm{L}$  and that there is a lot of freedom in experimental tunability of, e.g., the central dot level $\varepsilon_2$.
In particular, the lower right plot in Fig.~\hyperref[fig:fig2]{2(c)} shows that it is not required for the energy levels $\varepsilon_j$ of the quantum dots to have the same magnitude with alternating sign in order to obtain nontrivial topological phases.
For $w=2 \, \Gamma_\mathrm{L}$ and $\Gamma_\mathrm{NL} =0.8 \, \Gamma_\mathrm{L}$, $\varepsilon_2$ can be detuned even up to $\varepsilon_2 \approx - 13 \, \Gamma_\mathrm{L}$ with topological regions still being present. 

\section{\label{sec:geometryAndTopology}Quantum geometry and microwave spectroscopy}%
In the following, we aim at applying the method of microwave spectroscopy described in Ref.~\cite{Klees:2020} in the presence of more than one nondegenerate pair of ABS.
The single-particle QGT in Eq.~\eqref{eq:QGTnew} of each ABS $|\psi_{n}\rangle$ can be decomposed into the sum
\begin{align}
	q^{(n)}_{\alpha\beta} = 
	\sum_{m = -N  \atop m \neq 0,n}^N q_{\alpha\beta}^{(nm)}
	\label{eq:QGT}
\end{align}
of nonadiabatic transition elements $q_{\alpha\beta}^{(nm)} = \langle \partial_\alpha \psi_n|\psi_m\rangle \langle\psi_m|\partial_\beta\psi_n\rangle$.

\begin{figure}
	\centering
	\includegraphics[width=0.7\linewidth]{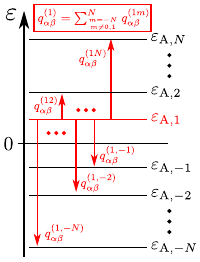}
	\caption{
		Different contributions of the elements $q_{\alpha\beta}^{(1m)}$ from other ABS with energies $\varepsilon_{\mathrm{A},m}$ $(m \neq 0,1)$ to the QGT $q_{\alpha\beta}^{(1)}$ of the ABS with energy $\varepsilon_{\mathrm{A},1}$ according to Eq.~\eqref{eq:QGT}.}
	\label{fig:fig3}
\end{figure}

In general, Eq.~\eqref{eq:QGT} implies a priori that we need to measure all possible nonadiabatic transitions between a certain state $| \psi_n \rangle$ and all the other states $| \psi_m \rangle$ $(n \neq m)$ by microwave spectroscopy  to construct the QGT of this particular single-particle state $| \psi_n \rangle$, as sketched in Fig.~\ref{fig:fig3}.
We will first describe the general procedure and later discuss possible experimental difficulties.

On the one hand, in order to measure the diagonal elements $q_{\alpha\alpha}^{(nm)}$ of the QGT, we need to apply a small perturbation $(A/\hbar \omega \ll 1)$ to one phase $\varphi_\alpha \to \varphi_\alpha + 2 A \cos(\omega t)/ \hbar \omega$, resulting in the transition (photon absorption) with a rate $R_{n \to m,\alpha\alpha} = r_{n\to m,\alpha\alpha}\,  \delta(\varepsilon_m - \varepsilon_n - \hbar \omega)$ with the oscillator strength \cite{Ozawa:2018ky,Klees:2020}
\begin{align}
	r_{n \to m,\alpha\alpha} 
	= 
	\frac{2\pi }{ \hbar } A^2 \, q_{\alpha\alpha}^{(nm)}
	\label{eq:oscillatorStrenghtDiagonal}
\end{align}
according to Fermi's golden rule. 
On the other hand, we assume to simultaneously apply two small perturbations, $\varphi_\alpha \to \varphi_\alpha + 2 A \cos(\omega t)/ \hbar \omega$ to one phase and $\varphi_\beta \to \varphi_\beta + 2 A \cos(\omega t- \gamma)/ \hbar \omega$ to another phase $(\alpha \neq \beta)$ with a fixed phase difference $\gamma$ between the two modulations, to measure the off-diagonal elements $q_{\alpha\beta}^{(nm)}$ of the QGT.
The resulting transition rates are $R_{n \to m,\alpha\beta}^{(\gamma)} = r_{n\to m,\alpha\beta}^{(\gamma)} \,  \delta(\varepsilon_m - \varepsilon_n - \hbar \omega)$ with the oscillator strength \cite{Ozawa:2018ky,Klees:2020}
\begin{multline}
	r_{n \to m,\alpha\beta}^{(\gamma)} = \frac{2\pi  A^2 }{\hbar}
	\Bigl\{ q_{\alpha\alpha}^{(nm)} + q_{\beta\beta}^{(nm)} \\
	+ e^{i\gamma}q_{\alpha\beta}^{(nm)} + e^{- i \gamma}q_{\beta\alpha}^{(nm)}  \Bigr\} ,
	\label{eq:oscillatorStrenghtOffdiagonal}
\end{multline}
which now depends on the relative phase difference $\gamma$ between the two drives.
Details of the derivation of Eqs.~\eqref{eq:oscillatorStrenghtDiagonal} and \eqref{eq:oscillatorStrenghtOffdiagonal} are presented in Appendix~\ref{app2:PMS}.
By comparing Eqs.~\eqref{eq:oscillatorStrenghtDiagonal} and \eqref{eq:oscillatorStrenghtOffdiagonal} with the definition of the QGT in Eq.~\eqref{eq:QGT}, we obtain the relations
\begin{subequations}
\label{eq:quantumGeometryRates}
\begin{align}
	g_{\alpha\alpha}^{(n)}
	&=
	\frac{ \hbar }{2\pi A^2} \sum_{m = -N  \atop m \neq 0,n}^N r_{n \to m,\alpha\alpha}  ,
	\\
	g_{\alpha\beta}^{(n)} &= \frac{\hbar} {8\pi  A^2 }\sum_{m = -N  \atop m \neq 0,n}^N 
	\bigl[  r_{n \to m,\alpha\beta}^{(0)} - r_{n \to m,\alpha\beta}^{(\pi)} \bigr] ,
	\\
	f_{\alpha\beta}^{(n)}  
	&=
	\frac{\hbar}{4\pi  A^2 }
	\sum_{m = -N  \atop m \neq 0,n}^N \bigl[ r_{n \to m,\alpha\beta}^{(\pi/2)} - r_{n \to m,\alpha\beta}^{(-\pi/2)} \bigr] .
\end{align}
\end{subequations}
This allows us to measure the metric tensor by linear microwave spectroscopy, i.e., modulating one phase or two phases with $\gamma = 0,\pi$, while the Berry curvature follows from circular microwave spectroscopy, i.e., modulating two phases with $\gamma = \pm\pi/2$.

Although this seems to be straightforward to implement, the experimental difficulty at very low temperature is to measure the transitions between two occupied or two empty states, respectively.
To overcome this obstacle, one could either perform the experiment at sufficiently high temperatures, for which previously filled (empty) states get partially depopulated (populated), or one could use pulses to empty/fill a certain state before measuring the transition. 

\begin{figure}
	\centering
	\includegraphics[width=\linewidth]{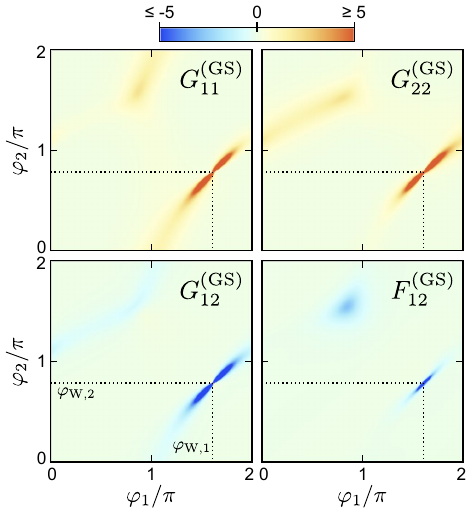}
	\caption{
		Ground state metric tensor $G_{12}^{\mathrm{(GS)}}$ and Berry curvature $F_{12}^{(\mathrm{GS})}$ for $N=3$ dots at $\varphi_3 = 0.4\, \pi$ in the topological phase with $C_{12}^{\mathrm{(GS)}} = -1$ [cf.~Fig.~\hyperref[fig:fig2]{2(c)}]. 
		The elements of the QGT are strongly peaked for values of $\boldsymbol{\varphi}$ close to a Weyl point. 
		The coordinates of the Weyl point indicated in this plot are $\boldsymbol{\varphi}_{\mathrm{W}} = (\varphi_{\mathrm{W},1},\varphi_{\mathrm{W},2},\varphi_{\mathrm{W},3}) \approx (1.603,  0.781, 0.384) \pi$ [cf.~lower red dot in Fig.~\hyperref[fig:fig2]{2(b)}].
		The visible range of values of the elements of the QGT has been limited from $-5$ to $+5$ to make the overall structure for small values visible. 
		In fact, the values of the elements of the QGT at the point $(\varphi_{\mathrm{W},1},\varphi_{\mathrm{W},2})$ are exceeding $\pm 800$.
		Parameters: $\Gamma_\mathrm{NL} = 0.8 \, \Gamma_\mathrm{L}$, $w = 0.2 \, \Gamma_\mathrm{L}$, $\varepsilon_1 = - \varepsilon_2 = \varepsilon_3 = - \Gamma_\mathrm{L}/2$, $\varphi_0 = 0$.}
	\label{fig:fig4}
\end{figure}

Fortunately, for low-temperature quantum transport, the experimentally relevant quantity is the QGT 
of the GS $Q^{\mathrm{(GS)}}_{\alpha\beta} = G^{\mathrm{(GS)}}_{\alpha\beta} - i F^{\mathrm{(GS)}}_{\alpha\beta}/2$ which, in the noninteracting case, can be constructed from the elements in Eq.~\eqref{eq:quantumGeometryRates} by means of Eq.~\eqref{eq:groundStateQGT}.
The resulting Berry curvature and metric tensor of the GS, $F_{\alpha\beta}^{\mathrm{(GS)}}$ and $G_{\alpha\beta}^{\mathrm{(GS)}}$, respectively, are still accessible via microwave spectroscopy even at zero temperature and the result simply reads (see Appendix~\ref{app2:PMS} for details)
\begin{subequations}
	\label{eq:quantumGeometryRatesGS}
	\begin{align}
		G_{\alpha\alpha}^{\mathrm{(GS)}} 
		&= 
		\frac{ \hbar }{2\pi A^2 } \sum_{j \in D^-} 
		\sum_{ k \in D^+}  r_{j \to k,\alpha\alpha} ,
		\label{eq:quantumGeometryRatesGSGaa}
		\\
		G_{\alpha\beta}^{\mathrm{(GS)}} 
		&=
		\frac{\hbar}{8 \pi  A^2 }  \sum_{j \in D^-} 
		\sum_{ k \in D^+} (r_{j\to k,\alpha\beta}^{(0)} -
		r_{j\to k,\alpha\beta}^{(\pi)}) , 
		\label{eq:quantumGeometryRatesGSGab}
		\\
		F_{\alpha\beta}^{\mathrm{(GS)}} 
		&= 
		\frac{\hbar}{4\pi A^2 }  \sum_{j \in D^-} 
		\sum_{ k \in D^+}
		\bigl( r_{j\to k,\alpha\beta}^{(\pi/2)} -
		r_{j\to k,\alpha\beta}^{(-\pi/2)} \bigr) ,
		\label{eq:quantumGeometryRatesGSFab}
	\end{align}
\end{subequations}
which shows that, indeed, only transitions from occupied $(D^-)$ to empty states $(D^+)$ are needed.
Both the GS Berry curvature and the GS metric tensor can therefore be simply obtained from all the experimentally accessible transitions by using microwave spectroscopy.
In Fig.~\ref{fig:fig4}, we show, as an example of Eq.~\eqref{eq:quantumGeometryRatesGS}, the metric tensor and the Berry curvature of the GS for $N=3$ dots in the topological region for $\varphi_3 = 0.4 \, \pi $ in which we find a Chern number $C_{12}^{\mathrm{(GS)}} = -1$ [cf. Fig.~\hyperref[fig:fig2]{2(c)}].
A local measurement of the elements of the QGT by microwave spectroscopy will not only reveal the topological phase in terms of the Chern number by integrating the Berry curvature, it is also possible to locate Weyl points in phase space $\mathbb{T}^N$ for which zero-energy states are expected.
In fact, Fig.~\ref{fig:fig4} shows that there are regions where the elements of the QGT are significantly different from zero, signaling the location of a Weyl point located at $\boldsymbol{\varphi}_{\mathrm{W}}/\pi \approx (1.603,  0.781, 0.384)$, with the $\varphi_1$ and $\varphi_2$ coordinates indicated in the plots.
A hint to another Weyl point is visible as a blurred point at $(\varphi_1, \varphi_2) \approx (0.9,1.5) \pi$ which is actually in a different $\varphi_3 = \mathrm{const.}$ plane [cf. Fig.~\hyperref[fig:fig2]{2(b)}] and all of the Weyl points can be located by many measurements for different fixed $\varphi_3$.
While possible zero-energy states might be useful on its own, let us recall here that the Chern number will be responsible for a quantized transconductance across two terminals \cite{Riwar:2016}.
\section{\label{sec:discussion}Discussion and conclusion}
In this work we have proposed an experimentally feasible system consisting of a linear chain of $N$ quantum dots connected to $N+1$ $s$-wave superconducting leads, as shown in Fig.~\ref{fig:fig1}, to study quantum geometry in topological Josephson matter.
This linear setup does not require a direct connection between the different superconducting terminals, as opposed to the system considered in Ref.~\cite{Klees:2020}, which might take a lot of engineering effort to realize in practice.
Furthermore, such quantum-dot models allow for the deterministic study of the local quantum geometry in topological Josephson matter in contrast to prior proposals based on statistical methods \cite{Riwar:2016,Eriksson:2017,Meyer:2017}.

We have derived an effective low-energy Hamiltonian of the chain, presented in Eq.~\eqref{eq:eq1}, which reveals large topological regions in terms of the integer-valued Chern number and, therefore, leaves large experimental freedom for the concrete values of the various system parameters to study quantum geometry and topology in the synthetic space of $N$ independent superconducting phases [cf.~Fig.~\ref{fig:fig2}]. 
In particular, as discussed in Fig.~\hyperref[fig:fig2]{2(c)}, it is also not necessary for the dots to be directly coupled $(w = 0)$ and it turns out that a larger value of the direct coupling can even lead to an enhancement of the robustness of the topological phase $(w = 2 \, \Gamma_\mathrm{L})$.
We furthermore believe that this or similar engineered systems also provide useful and easily scalable condensed-matter realizations to study higher-dimensional topological invariants such as, e.g., higher-dimensional Chern numbers \cite{Zhang:2001,Kraus:2013,Price:2015,Ozawa:2016,Chan:2017,Lohse:2018,Zilberberg:2018,Sugawa:2018,Lu:2018,Petrides:2018,Lee:2018,Weisbrich:2020}
or the so-called Dixmier-Douady invariants \cite{Murray:1996,Palumbo:2018,Palumbo:2019,Tan:2020,Chen:2020a}, since the synthetic dimension $N$ of the space of superconducting phases $\mathbb{T}^N$ scales with the number of superconducting leads which can be, in principle, arbitrary.

Moreover, we have worked out the insightful connection between the QGT of single-particle ABS and the QGT of the corresponding noninteracting GS.
At this point we want to stress that this connection is relevant to any physical system hosting multiple energy bands. 
As only the geometrical property of the GS is of particular importance for low-temperature experiments, we have shown that, although the QGT of an individual ABS cannot be accessed without further preparation, the QGT of the GS as a whole as presented in Eq.~\eqref{eq:quantumGeometryRatesGS} is still accessible by microwave spectroscopy from the oscillator strengths of multiple absorption lines.
In this regard, we have generalized the previously discussed application of microwave spectroscopy to measure the QGT of topological Josephson matter in Ref.~\cite{Klees:2020} to the case of the presence of multiple ABS.
We emphasize that this method provides an alternative way to measure the GS topology beyond the already mentioned transconductance measurements originally proposed in Ref.~\cite{Riwar:2016}.
Furthermore, microwave spectroscopy provides a tool to gain access to the local QGT and makes it possible to localize the position of zero-energy states through the measurement of Weyl points in phase space [cf.~Fig.~\ref{fig:fig4}].

Since it is based on quantum dots, our system provides an ideal starting point to study the effect of Coulomb interactions on the geometry and topology of the GS.
Futhermore, it may be relevant for concrete experiments to study the effect of a finite superconducting gap, which can be of the order of the tunneling \cite{Hofstetter:2009,Herrmann:2010}, and the influence of the continuum \cite{Zgirski:2011,Repin:2019} on the quantum geometry of the GS, which is not captured in our low-energy theory at infinite gap.
Finally, the general role of a finite temperature and dissipation on both quantum geometry and topology remains an open question, both theoretically and experimentally \cite{Dittmann:1992,Viyuela:2014,Viyuela:2018,Gneiting2020}, and it will be interesting to apply these concepts to the proposed system in the future.
\begin{acknowledgments}
	We thank Hannes Weisbrich, Hugues Pothier, and Cristian Urbina for useful discussions. 
	This work was supported by the Deutsche Forschungsgemeinschaft through SFB 767 and Grant No. RA 2810/1. 
	J.C.C. acknowledges the support via the Mercator Program of the Deutsche Forschungsgemeinschaft in the frame of the SFB 767.
\end{acknowledgments}
\appendix
\section{\label{app1:effHam}Effective Hamiltonian for $N$ dots}
Here, we derive the effective Hamiltonian of $N$ dots connected to $N+1$ superconducting terminals as sketched in Fig.~\ref{fig:fig1}. 
Due to the inclusion of superconductivity, we start by defining a set of Pauli matrices ($\tau_1$, $\tau_2$, $\tau_3$) in Nambu space.
The starting Hamiltonian reads
\begin{align}
	H = H_\mathrm{D} +  H_\mathrm{S} +  H_\mathrm{DS} .
\end{align}
The $N$ dots and their couplings are described by $H_\mathrm{D} = \boldsymbol{d}^{\dag} \hat{H}_\mathrm{D} \boldsymbol{d}^{\phantom\dag}$, where the $2N \times 2N$ matrix Hamiltonian is given by
\begin{align}
	 \hat{H}_\mathrm{D} = \begin{pmatrix}
	 \varepsilon_1  & w_{12} & 0 & \cdots & 0 & 0 \\
	 w_{12} & \varepsilon_2  & w_{23} & \cdots & 0 & 0 \\
	 0 & w_{23} & \varepsilon_3  & \cdots & 0 & 0 \\
	 \vdots & \vdots & \vdots & \ddots & \vdots & \vdots \\
	 0 & 0 & 0 & \cdots & \varepsilon_{N-1}  & w_{N-1,N} \\
	 0 & 0 & 0 & \cdots & w_{N-1,N} & \varepsilon_{N}  \\
	 \end{pmatrix}
	 \otimes \tau_3
	\label{eq:dots}
\end{align}
with $\varepsilon_j$ being the onsite energy of dot $j$, $w_{j,j+1}$ being the hopping energy between dots $j$ and $j+1$, and $\boldsymbol{d}^{\dag} = (d_{1\uparrow}^{\dag},d_{1\downarrow}^{\phantom\dag}, d_{2\uparrow}^{\dag},d_{2\downarrow}^{\phantom\dag}, \ldots, d_{N\uparrow}^{\dag},d_{N\downarrow}^{\phantom\dag})$ is the Nambu spinor consisting of electronic annihilation (creation) operators $d_{j\sigma}^{(\dag)}$ of spin $\sigma = \uparrow,\downarrow$ on dot $j$.

The Hamiltonian describing the $N+1$ BCS superconducting leads reads $H_{\mathrm{S}} 
= \sum_{\boldsymbol{k} }  \boldsymbol{c}_{\boldsymbol{k}}^{\dag} \hat{H}_{\mathrm{S},\boldsymbol{k}} \boldsymbol{c}_{\boldsymbol{k}}^{\phantom\dag}$, with $\boldsymbol{c}_{\boldsymbol{k}}^{\dag} = (c_{0\boldsymbol{k}\uparrow}^{\dag},c_{0(-\boldsymbol{k})\downarrow}^{\phantom\dag}, c_{1\boldsymbol{k}\uparrow}^{\dag},c_{1(-\boldsymbol{k})\downarrow}^{\phantom\dag}, \ldots, c_{N\boldsymbol{k}\uparrow}^{\dag},c_{N(-\boldsymbol{k})\downarrow}^{\phantom\dag})$ containing electronic annihilation (creation) operators $c_{j\boldsymbol{k}\sigma}^{(\dag)}$ of quasimomentum $\boldsymbol{k}$ and spin $\sigma$ in lead $j$, and the $2(N+1) \times 2(N+1)$ block-diagonal matrix Hamiltonian
\begin{align}
	\hat{H}_{\mathrm{S},\boldsymbol{k}} = \mathrm{diag}(\hat{H}_{\mathrm{S},0\boldsymbol{k}},\hat{H}_{\mathrm{S},1\boldsymbol{k}},\ldots,\hat{H}_{\mathrm{S},N\boldsymbol{k}}) .
	\label{eq:superconductors}
\end{align}
Here, $\hat{H}_{\mathrm{S},j\boldsymbol{k}} =  \xi_{j\boldsymbol{k}}  \tau_3 + \Delta_j e^{i \varphi_j \tau_3} \tau_1$, where $\Delta_j$ and $\varphi_j$ are the superconducting pairing potential and phase, respectively, and $\xi_{j\boldsymbol{k}}$ is the normal state dispersion in each lead.

Finally, the coupling between the dots and neighboring superconducting leads is given by
\begin{align}
	H_{\mathrm{DS}} 
	&=  \sum_{\boldsymbol{k} } ( \boldsymbol{d}^{\dag} \hat{V}_\mathrm{DS} \boldsymbol{c}_{\boldsymbol{k}}^{\phantom\dag} + \mathrm{h.c.})  ,
	\label{eq:tunnelingDS}
\end{align}
with the $2N \times 2(N+1)$ matrix coupling 
\begin{align}
	\hat{V}_\mathrm{DS} =
	\begin{pmatrix}
		t_{10}  & t_{11}  & 0 & 0 & \cdots & 0 & 0 \\
		0 & t_{21}  & t_{22}   & 0  & \cdots & 0 & 0 \\
		0 & 0 & t_{32}  & t_{33}   & \cdots & 0 & 0 \\
		\vdots & \vdots & \vdots & \vdots & \ddots & \vdots & \vdots \\
		0 & 0 & 0 & 0  & \cdots & t_{N,N-1}  & t_{NN}  \\
	\end{pmatrix}
	\otimes \tau_3 
\end{align}
and the hopping energy $t_{jk}$ between dot $j$ and lead $k$. 

We use the Green's function (GF) technique to derive an effective Hamiltonian of the subsystem of the dots. From $\hat{g}_{\mathrm{S},j\boldsymbol{k}} = (\varepsilon - \hat{H}_{\mathrm{S},j\boldsymbol{k}})^{-1}$, we find the $2 \times 2$ GF of the $j$-th lead as 
\begin{align}
	\hat{g}_{\mathrm{S},j\boldsymbol{k}} = \frac{\varepsilon \, + \xi_{j\boldsymbol{k}} \tau_3 + \Delta_j e^{i\varphi_j \tau_3}\tau_1 }{\varepsilon^2 - \xi_{j\boldsymbol{k}}^2 - \Delta_j^2 } .
\end{align}
Since the dots and hoppings do not depend on quasimomentum $\boldsymbol{k}$, we sum over all quasimomenta and define $\hat{g}_{\mathrm{S},j} = \sum_{\boldsymbol{k} } \hat{g}_{\mathrm{S},j\boldsymbol{k}}$. 
Turning the summation into an energy integration in the wide-band limit via $\sum_{\boldsymbol{k} } \to N_0 \int \mathrm{d}\xi_{j\boldsymbol{k}}$, with $N_0$ being the normal state density of states at the Fermi energy, we obtain
\begin{align}
	\hat{g}_{\mathrm{S},j} = - \pi N_0 \frac{\varepsilon + \Delta_j e^{i\varphi_j \tau_3}\tau_1 }{\sqrt{ \Delta_j^2- \varepsilon^2 } }
	\stackrel{\Delta_j \to \infty}{\longrightarrow}  - \pi N_0 e^{i\varphi_j \tau_3}\tau_1 ,
\end{align}
where we assume that the pairing potentials are larger than all relevant energy scales. 
Finally, the bare $2(N+1) \times 2(N+1)$ matrix GF of the subsystem of the superconducting leads is given by the block-diagonal matrix 
\begin{align}
\hat{g}_\mathrm{S} = \mathrm{diag}(\hat{g}_{\mathrm{S},0}, \ldots , \hat{g}_{\mathrm{S},N}),
\end{align}
written in the basis $\boldsymbol{c}^{\dag} = (\boldsymbol{c}_{0}^{\dag}, \ldots , \boldsymbol{c}_{N}^{\dag}) = \sum_{\boldsymbol{k} } \boldsymbol{c}_{\boldsymbol{k}}^{\dag}$.

Formally, the bare $2N \times 2N$ matrix GF of the dots with their couplings is given by $\hat{g}_\mathrm{D} = (\varepsilon - \hat{H}_\mathrm{D})^{-1}$. 
The explicit form of $\hat{g}_\mathrm{D}$ is not needed for the following derivation.

We define an effective matrix Hamiltonian $\hat{H}_\mathrm{eff}^{(N)}$ from the dressed GF of the $2N\times 2N$ dot subsystem, i.e., $\hat{G}_\mathrm{D} = (\varepsilon - \hat{H}_\mathrm{eff}^{(N)})^{-1}$, which is the solution of the Dyson equation $\hat{G}_\mathrm{D} = \hat{g}_\mathrm{D} + \hat{g}_\mathrm{D} \hat{\Sigma}_\mathrm{D} \hat{G}_\mathrm{D}$, with the self-energy $\hat{\Sigma}_\mathrm{D} = \hat{V}_\mathrm{DS} \,  \hat{g}_\mathrm{S} \, \hat{V}_\mathrm{DS}^\dag$. 
Using the formal definition of $\hat{g}_\mathrm{D}$, we obtain the solution $\hat{H}_\mathrm{eff}^{(N)} =  \hat{H}_\mathrm{D} + \hat{\Sigma}_\mathrm{D}$ which describes the dots proximity coupled to the superconductors. 
Finally, the effective low-energy Hamiltonian is
\begin{align}
	H_\mathrm{eff} &=  \boldsymbol{d}^\dag \hat{H}_\mathrm{eff}^{(N)} \boldsymbol{d}
	= H_\mathrm{D} + H_\mathrm{LP} + H_\mathrm{NLP} ,
	\label{eq:effectiveHarbitraryN}
\end{align}
with the local and nonlocal pairings (LP and NLP) originating from the self-energy $\hat{\Sigma}_\mathrm{D}$ given by 
\begin{subequations}
	\label{eq:pairing}
	\begin{align}
		H_{\mathrm{LP}} &= \sum_{j = 1}^{N}  \boldsymbol{d}_j^{\dag}  
		\bigl(  \Gamma_{j}^{(j-1)} e^{i \varphi_{j-1} \tau_3} 
		+
		\Gamma_{j}^{(j)}   e^{i \varphi_j \tau_3} \bigr) 
		\tau_1 \boldsymbol{d}_{j}^{\phantom\dag}  ,
		\\
		H_{\mathrm{NLP}} &= \sum_{j=1}^{N-1} \Gamma_{j,j+1}^{(j)} \, ( \boldsymbol{d}_j^{\dag} e^{i \varphi_j \tau_3}\tau_1 \boldsymbol{d}_{j+1}^{\phantom\dag} 
		+
		\boldsymbol{d}_{j+1}^{\dag} e^{i \varphi_j \tau_3}\tau_1 \boldsymbol{d}_{j}^{\phantom\dag} )	,
	\end{align}
\end{subequations}
with $\Gamma_{j}^{(k)} = \pi N_0 t_{jk}^2$, $\Gamma_{j , j+1}^{(k)} = \pi N_0 t_{jk} t_{j+1,k}$ and the local Nambu spinors $\boldsymbol{d}_j^\dag = (d_{j\uparrow}^\dag,d_{j\downarrow}^{\phantom\dag})$.
Eq.~\eqref{eq:effectiveHarbitraryN}, together with Eq.~\eqref{eq:pairing}, represents the effective Hamiltonian presented in Eq.~\eqref{eq:eq1} in Sec.~\ref{sec:model} of the main text.

\section{\label{app3:manybody}Quantum geometry of general fermionic $n$-particle states}
For $n \in \mathbb{N}$, let $| \psi_j \rangle$, $j \in \{1,\ldots,n\}$, be single-particle states with their individual single-particle quantum geometric tensors (QGTs)
\begin{align}
	q^{(j)}_{\alpha\beta} &= 
	\langle \partial_\alpha \psi_{j} |( 1 - |\psi_{j} \rangle\langle\psi_{j} |) |\partial_\beta\psi_{j}\rangle ,
\end{align}
their Berry curvatures $f^{(j)}_{\alpha\beta} = -2 \, \mathrm{Im}(q^{(j)}_{\alpha\beta})$ and their (Fubini-Study) metric tensors $g^{(j)}_{\alpha\beta} = \mathrm{Re}(q^{(j)}_{\alpha\beta})$. 
The different single-particle states shall be orthonormalized, i.e., $\langle \psi_j | \psi_k \rangle = \delta_{jk}$.
Furthermore, let 
\begin{align}
	| \Psi_{n} \rangle = \frac{1}{\sqrt{n!}} \sum_{\mathcal{P}} (-1)^{p(\mathcal{P})} \, \mathcal{P}\Bigl[ | \psi_{1} \rangle \otimes \cdots \otimes | \psi_{n} \rangle \Bigr] 
	\label{eq:manybodystate}
\end{align}
be the corresponding general antisymmetric noninteracting $n$-particle fermionic state, in which we sum over all $n!$ different permutations with the permutation operator $\mathcal{P}$ and the number of transpositions $p(\mathcal{P})$ of a particular permutation.
How are the single-particle QGTs related to the corresponding noninteracting $n$-particle QGT
\begin{align}
	Q^{(n)}_{\alpha\beta} &= 
	\langle \partial_\alpha \Psi_{n} |( 1 - |\Psi_{n} \rangle\langle\Psi_{n} |) |\partial_\beta\Psi_{n}\rangle
\end{align}
and, in particular, what are the corresponding relations to the Berry curvature $F^{(n)}_{\alpha\beta} = -2 \, \mathrm{Im}(Q^{(n)}_{\alpha\beta})$ and the metric tensor $G^{(n)}_{\alpha\beta} = \mathrm{Re}(Q^{(n)}_{\alpha\beta})$ of the $n$-particle state $| \Psi_{n} \rangle$?

After a lengthy but straightforward calculation, the final result reads
\begin{subequations}
	\label{eq:manybodyresult}
\begin{align}
	G^{(n)}_{\alpha\beta} 
	&= \sum_{j = 1}^n g_{\alpha\beta}^{(j)} 
		- \sum_{j,k =1 \atop j \neq k}^n  
			\langle \partial_{\alpha} \psi_j | \psi_k \rangle
			\langle \psi_k | \partial_{\beta} \psi_j \rangle \, ,
			\label{eq:manybodyresult1}
\\
	F^{(n)}_{\alpha\beta} 
	&= \sum_{j = 1}^n f_{\alpha\beta}^{(j)} \, ,
	\label{eq:manybodyresult2}
\end{align}
\end{subequations}
which is also presented in Eq.~\eqref{eq:groundStateQGT} in Sec.~\ref{sec:model} of the main text for the GS of the system that contains $N$ occupied single-particle ABS.
While the result for the metric tensor reveals a nontrivial and less intuitive relation, the result for the Berry curvature can be understood more intuitively and matches the expected result. 
On an intuitive level, suppose that the single-particle states acquire the Berry phases $\phi_j$ according to $| \psi_j \rangle \mapsto e^{i \phi_j} | \psi_j \rangle$ after a cyclic evolution in parameter space \cite{Berry:1984}. 
Then, the evolution of the corresponding $n$-particle state results in 
\begin{align}
	| \Psi_{n} \rangle \mapsto e^{i \Phi_n} | \Psi_{n} \rangle
\end{align}
according to the definition in Eq.~\eqref{eq:manybodystate}, where the Berry phase $\Phi_n = \sum_{j = 1}^n \phi_j$ is the sum of the individual Berry phases $\phi_j$. 
This directly translates to the Berry curvature by means of Stokes' integral theorem. 

As a direct consequence of Eq.~\eqref{eq:manybodyresult2}, the Chern number of the GS is simply given by the sum of all individual Chern numbers of the $N$ occupied single-particle states as expected, i.e.,
\begin{align}
C_{\alpha\beta}^{\mathrm{(GS)}} = \sum_{j = -N }^{-1} c_{\alpha\beta}^{(j)} ,
\end{align}
which is also the result presented in Eq.~\eqref{eq:chernGS} in Sec.~\ref{sec:model} of the main text.

\section{\label{app2:PMS}Microwave spectroscopy of the quantum geometry of the ground state}
For the diagonal elements, $q_{\alpha\alpha}^{(n)} = g^{(n)}_{\alpha\alpha}$, we need to apply a perturbation to one phase $\varphi_\alpha \to \varphi_\alpha + 2 A \cos(\omega t)/ \hbar \omega$, resulting the transition rate $R_{n \to m,\alpha\alpha} = r_{n \to m,\alpha\alpha}\,  \delta(\varepsilon_m - \varepsilon_n - \hbar \omega)$ with the oscillator strength \cite{Ozawa:2018ky,Klees:2020}
\begin{align}
	r_{n \to m,\alpha\alpha} = \frac{2\pi A^2 \bigl| \langle \psi_m | (\partial_\alpha \hat{H}_{\mathrm{eff}}^{(N)})  | \psi_n \rangle \bigr|^2 }{ \hbar (\varepsilon_m - \varepsilon_n)^2}
\end{align}
according to Fermi's golden rule. 
In order to measure the off-diagonal elements of the QGT $q^{(n)}_{\alpha\beta}$ for $\alpha \neq \beta$, we need to simultaneously apply the perturbations $\varphi_\alpha \to \varphi_\alpha + 2 A \cos(\omega t)/ \hbar \omega$ and $\varphi_\beta \to \varphi_\beta + 2 A \cos(\omega t - \gamma)/ \hbar \omega$ with a phase difference $\gamma$, resulting in the transition rates $R_{n \to m,\alpha\beta}^{(\gamma)} = r_{n \to m,\alpha\beta}^{(\gamma)} \, \delta(\varepsilon_m - \varepsilon_n - \hbar \omega)$ with the oscillator strength \cite{Ozawa:2018ky,Klees:2020}
\begin{align}
	r_{n\to m,\alpha\beta}^{(\gamma)} = \frac{2\pi A^2 \bigl| \langle \psi_m | \bigl( \partial_\alpha \hat{H}_{\mathrm{eff}}^{(N)} 
		+ e^{i \gamma} \partial_\beta \hat{H}_{\mathrm{eff}}^{(N)} \bigr) | \psi_n \rangle \bigr|^2}{\hbar (\varepsilon_m - \varepsilon_n)^2} 
	\label{eq:rofgammaDefinition}
\end{align}
according to Fermi's golden rule. 
Note that $r_{n \to m,\alpha\beta}^{(\gamma)} = r_{m\to n,\alpha\beta}^{(-\gamma)}$ by definition. 
Using the identities
\begin{subequations}
\begin{align}
	\langle \partial_\alpha \psi_n |  \psi_m \rangle &= -  \langle  \psi_n |  \partial_\alpha \psi_m \rangle ,
	\\
	\langle \psi_n | (\partial_\alpha \hat{H}_{\mathrm{eff}}^{(N)}) | \psi_m \rangle &= (\varepsilon_n - \varepsilon_m) \langle \partial_\alpha \psi_n |  \psi_m \rangle ,
\end{align}
\end{subequations}
for $n \neq m$, we obtain
\begin{subequations}
\begin{align}
	r_{n \to m,\alpha\alpha} &= \frac{2\pi A^2 }{ \hbar } q_{\alpha\alpha}^{(nm)} ,
	\label{eq:rnotofgamma}
	\\
	r_{n\to m,\alpha\beta}^{(\gamma)} &= \frac{2\pi  A^2 }{\hbar}
	\Bigl\{ q_{\alpha\alpha}^{(nm)} + q_{\beta\beta}^{(nm)} 
	\nonumber \\
	&\qquad \qquad \quad + e^{i\gamma}q_{\alpha\beta}^{(nm)} + e^{- i \gamma}q_{\beta\alpha}^{(nm)}  \Bigr\} .
	\label{eq:rofgamma}
\end{align}
\end{subequations}
which are exactly Eqs.~\eqref{eq:oscillatorStrenghtDiagonal} and \eqref{eq:oscillatorStrenghtOffdiagonal} as presented in Sec.~\ref{sec:geometryAndTopology} of the main text.
Finally, for $\alpha \neq \beta$, this yields the useful identities
\begin{subequations}
\begin{align}
	\mathrm{Re}(q_{\alpha\beta}^{(nm)})  
	&=
	\frac{\hbar}{8 \pi  A^2 } (r_{n\to m,\alpha\beta}^{(0)} -
	r_{n\to m,\alpha\beta}^{(\pi)}) ,
	\label{eq:realPartQGT}
\\
	\mathrm{Im}( q_{\alpha\beta}^{(nm)} )
	&=
	-  \frac{\hbar}{8\pi A^2 } \bigl( r_{n\to m,\alpha\beta}^{(\pi/2)} -
	r_{n\to m,\alpha\beta}^{(-\pi/2)} \bigr) .
	\label{eq:imaginaryPartQGT}
\end{align}
\end{subequations}
At zero temperature, the GS is given by an (antisymmetric) combination of all $N$ occupied single-particle states up to the Fermi energy, as generally defined in Eq.~\eqref{eq:manybodystate}.
We now use the general result in Eq.~\eqref{eq:manybodyresult} to relate the measured oscillator strengths from microwave spectroscopy to the Berry curvature $F_{\alpha\beta}^{\mathrm{(GS)}}$ and the metric tensor  $G_{\alpha\beta}^{\mathrm{(GS)}}$ of the ground state.

For the Berry curvature, we obtain
\begin{widetext}
	\begin{align}
		&F_{\alpha\beta}^{\mathrm{(GS)}} 
		\stackrel{\eqref{eq:manybodyresult2}}{=}
		\sum_{j=-N}^{-1}
		f_{\alpha\beta}^{(j)} 
		\stackrel{\eqref{eq:QGT}}{=} 
		-2 \sum_{j=-N}^{-1}  \sum_{k = -N  \atop k \neq 0,j}^N \mathrm{Im} (q_{\alpha\beta}^{(jk)})
		\stackrel{\eqref{eq:imaginaryPartQGT}}{=} 
		\frac{\hbar}{4\pi A^2 }  \sum_{j=-N}^{-1}  \sum_{k = -N  \atop k \neq 0,j}^N 
		\bigl( r_{j\to k,\alpha\beta}^{(\pi/2)} -
		r_{j\to k,\alpha\beta}^{(-\pi/2)} \bigr)
		\nonumber \\
		&= 
		\frac{\hbar}{4\pi A^2 } \Biggl[ \underbrace{ \sum_{j,k =-N \atop k \neq j}^{-1} 
			\bigl( r_{j\to k,\alpha\beta}^{(\pi/2)} -
			r_{j\to k,\alpha\beta}^{(-\pi/2)} \bigr) 
		}_{ =\,  0, \text{ Eq.}\,\eqref{eq:rofgammaDefinition} }
		+ 
		\sum_{j=-N}^{-1}  \sum_{k = 1}^N 
		\bigl( r_{j\to k,\alpha\beta}^{(\pi/2)} -
		r_{j\to k,\alpha\beta}^{(-\pi/2)} \bigr) \Biggr]
		= 
		\frac{\hbar}{4\pi A^2 }  \sum_{j=-N}^{-1}  \sum_{k = 1}^N 
		\bigl( r_{j\to k,\alpha\beta}^{(\pi/2)} -
		r_{j\to k,\alpha\beta}^{(-\pi/2)} \bigr) \, ,
	\end{align}
which is the one presented in Eq.~\eqref{eq:quantumGeometryRatesGSFab} in Sec.~\ref{sec:geometryAndTopology} of the main text. 
For the metric tensor, we obtain
\begin{align}
	G_{\alpha\beta}^{\mathrm{(GS)}} 
	&\stackrel{\eqref{eq:manybodyresult1}}{=} \sum_{j = -N}^{-1} g_{\alpha\beta}^{(j)} 
	- \sum_{j,k = -N \atop j \neq k}^{-1}  
	\left\langle \partial_{\alpha} \psi_j \middle| \psi_k \right\rangle
	\left\langle \psi_k \middle| \partial_{\beta} \psi_j \right\rangle
	\stackrel{\eqref{eq:QGT}}{=}  
	\sum_{j = -N}^{-1} \sum_{k = -N  \atop k \neq 0,j}^N \mathrm{Re} (q_{\alpha\beta}^{(jk)})
		-  \sum_{j,k = -N \atop k \neq j}^{-1}  
		q_{\alpha\beta}^{(jk)}
	\nonumber \\
	&=
	\sum_{j = -N}^{-1} \sum_{k = -N  \atop k \neq 0,j}^N \mathrm{Re} (q_{\alpha\beta}^{(jk)})
	- \frac{1}{2} \sum_{j,k = -N \atop k \neq j}^{-1}  \underbrace{(q_{\alpha\beta}^{(jk)} + q_{\alpha\beta}^{(kj)})}_{2 \, \mathrm{Re}(q_{\alpha\beta}^{(jk)})}
	=
	\sum_{j = -N}^{-1} 
	\sum_{k = 1 }^N \mathrm{Re} (q_{\alpha\beta}^{(jk)})
	\, .
\end{align}
In both cases we see that we only need to measure transitions between occupied and empty states.
Finally, we get for the diagonal and off-diagonal elements
\begin{subequations}
\begin{align}
	G_{\alpha\alpha}^{\mathrm{(GS)}} 
	&\stackrel{\eqref{eq:rnotofgamma}}{=} 
	 \frac{ \hbar }{2\pi A^2 } \sum_{j = -N}^{-1} 
	\sum_{k = 1}^N  r_{j \to k,\alpha\alpha}  \, ,
	\\
	G_{\alpha\beta}^{\mathrm{(GS)}} 
	&\stackrel{\eqref{eq:realPartQGT}}{=} 
	\frac{\hbar}{8 \pi  A^2 }  \sum_{j = -N}^{-1} 
	\sum_{k = 1}^N (r_{j\to k,\alpha\beta}^{(0)} -
	r_{j\to k,\alpha\beta}^{(\pi)}) \, ,
\end{align}
\end{subequations}
which are the relations presented in Eqs.~\eqref{eq:quantumGeometryRatesGSGaa} and \eqref{eq:quantumGeometryRatesGSGab}, respectively, in \hyperref[sec:geometryAndTopology]{Sec.\,III} of the main text.
\end{widetext}

%
%

\end{document}